# Mean squared displacement and sinuosity of three-dimensional random search movements

Simon Benhamou

CEFE-CNRS, Montpellier, France

**Abstract.** Correlated random walks (CRW) have been used for a long time as a null model for animal's random search movement in two dimensions (2D). An increasing number of studies focus on animals' movement in three dimensions (3D), but the key properties of CRW, such as the way the mean squared displacement is related to the path length, are well known only in 1D and 2D. In this paper I derive such properties for 3D CRW, in a consistent way with the expression of these properties in 2D. This should allow 3D CRW to act as a null model when analysing actual 3D movements similarly to what is done in 2D.

Properties of Correlated Random Walks (CRW) in two-dimensional (2D) space, which are classically used to model 2D random search movements and thereby to serve as null model for more complex movements, are well known (Codling et al., 2008). This is not yet the case for CRW in three-dimensional (3D) space, whereas an increasing number of studies based on various recording techniques focus on 3D movements (Wilson et al., 2008; Voesenek et al., 2016; Le Bras et al., 2017; de Margerie et al., 2018). In particular, we are still missing mathematical expressions for two key metrics of CRWs in 3D space: the mean squared displacement (MSD), i.e. the expected value of the squared beeline distance from the starting point, and the path sinuosity, which corresponds to the spatial component of the diffusion coefficient (Benhamou, 2006). Here I show how these metrics can be expressed for 3D CRW, consistently with their expressions for 2D CRW.

When expressed in an exocentric (environment-bound) frame of reference ($X$, $Y$, $Z$), a 3D path starting at ($X_0$, $Y_0$, $Z_0$) can be modelled iteratively in a discrete way as

$X_i = X_{i-1} + l_i \cos(\phi_i) \cos(\theta_i)$

$Y_i = Y_{i-1} + l_i \cos(\phi_i) \sin(\theta_i)$

$Z_i = Z_{i-1} + l_i \sin(\phi_i)$

where $l_i$ is the length of the $i$th step, $\theta_i$ is the azimuth of this step, that is its orientation in the horizontal ($XY$) plane, and $\phi_i$ is its longitudinal inclination (restricted between $-\pi/2$ and $\pi/2$ radians), corresponding to the orientation in the vertical plane at azimuth $\theta_i$. Both azimuth $\theta_i$ and longitudinal inclination $\phi_i$ define the movement direction (i.e. heading). The lateral inclination, $\lambda_i$, which corresponds to the orientation in the frontal plane (i.e. orthogonally to the posterior-to-anterior axis), with respect to the horizontal level, does not appear here as it matters only when reorientations are expressed in an egocentric (animal-bound) frame of reference (see below). The squared displacement of any 3D discrete walk from initial location ($X_0$, $Y_0$, $Z_0$) to the location reached after $n$ steps ($X_n$, $Y_n$, $Z_n$) is therefore equal to:

$$\begin{aligned} R_{n,3D}^2 &= \left(\sum_{i=1}^{n} X_i - X_{i-1}\right)^2 + \left(\sum_{i=1}^{n} Y_i - Y_{i-1}\right)^2 + \left(\sum_{i=1}^{n} Z_i - Z_{i-1}\right)^2 \\ &= \sum_{i=1}^{n} l_i^2 + 2\sum_{i=1}^{n-1} \sum_{j=i+1}^{n} l_i l_j \sin(\phi_i)\sin(\phi_j) + \cos(\phi_i)\cos(\phi_j)\cos(\theta_j - \theta_i) \end{aligned} \quad (1)$$

In 2D paths, each reorientation is expressed as a single rotation $\alpha$ performed in the $XY$ plane, which also corresponds to the animal's coronal plane. In 3D paths, each reorientation can be expressed, in the egocentric frame of reference, as the result of three successive rotations: yaw, $\alpha$, in the coronal plane, i.e. around the heave (ventral-to-dorsal) axis, pitch, $\beta$, in the sagittal plane, i.e. around the sway (right-to-left) axis, and roll, $\rho$, in the frontal plane, i.e. around the surge (posterior-to-anterior) axis. Although $\rho$ does not directly affect the heading ($\theta$, $\phi$), it controls the lateral inclination $\lambda$, which determines to which extent $\alpha$ and $\beta$ are translated in terms of changes in θ and ϕ. For a given value of λ, the result of the combined action of α and β can be expressed as the so-called orthodromic arc (Fig. 1a), which is the smallest arc linking headings ($\theta_i$, $\phi_i$) and ($\theta_{i+1}$, $\phi_{i+1}$) (Fig. 1b). It can be characterized by its angular size $\omega_i$ (ranging between 0 and π radians), which corresponds to the angular deviation between these two headings, and by its initial and final orientations, $\delta_i$ and $\delta_i'$, measured in the frontal plane with respect to the horizontal level. Note that the change in arc orientation from $\delta_i$ to $\delta_i'$ leads to a passive change in lateral inclination from $\lambda_i$ to $\lambda_i'$ of the same magnitude ($\lambda_i' = \lambda_i + \delta_i' - \delta_i$), due to a rotation of the frontal plane by $\delta_i' - \delta_i$. The trigonometric relationships between reorientations expressed as (α, β) or as (ω, δ) and changes in azimuth θ and in longitudinal inclination ϕ are provided in the Appendix. Those relating such changes to (ω, δ) are well known in the context of great-circle navigation (e.g. Alerstam & Pettersson, 1991), where θ and ϕ correspond to the longitude and the latitude, respectively, δ corresponds to the initial movement direction with respect to East, and the shortest route length is given by the product of the arc size ω (in radians) by the Earth radius. The following relationship, which can be easily derived from the equations given in the Appendix or directly based on the spherical law of cosines, is worth mentioning here:

$$\cos(\Omega_{i,j}) = \sin(\phi_i)\sin(\phi_j) + \cos(\phi_i)\cos(\phi_j)\cos(\theta_i - \theta_j),$$

where $\Omega_{i,j}$ is the angular deviation between any two 3D headings ($\theta_i$, $\phi_i$) and ($\theta_j$, $\phi_j$) (note: $\Omega_{i,i+1} = \omega_i$). Eq. (1) can therefore be expressed as:

$$R_{n,3D}^2 = \sum_{i=1}^{n} l_i^2 + 2\sum_{i=1}^{n-1}\sum_{j=i+1}^{n} l_i l_j \cos(\Omega_{i,j}) \qquad (2)$$

Note that the term $l_i l_j \cos(\Omega_{i,j})$ corresponds to the scalar product $\mathbf{l}_i \cdot \mathbf{l}_j$, where $\mathbf{l}_i$ is a vector representing the $i^{th}$ step ($\mathbf{l}_i = \mathbf{X}_i - \mathbf{X}_{i-1}$, with $\mathbf{X}_i = (X_i, Y_i, Z_i)$).

In a 3D CRW, headings ($\theta$, $\phi$) are serially correlated, involving some persistence in the movement direction, but step lengths $l_i$ are not serially correlated and not cross-correlated with any angular variable, and yaw $\alpha$ and pitch $\beta$ are independent random variables. The general MSD formulation for 2D CRW (Kareiva & Shigesada, 1983; Benhamou, 2006), in which left and right turns are not necessarily balanced, is extremely complex, and eventually rarely used in practice. Its extension in 3D is beyond the scope of this paper, where I focus only on balanced 3D CRW, involving that $\alpha$ and $\beta$ distributions are both characterised by null mean sines ($s_\alpha = s_\beta = 0$) and concentration parameters equal to mean cosines $c_\alpha$ and $c_\beta$, respectively. Consequently, arc size $\omega$ (with $\cos(\omega) = \cos(\alpha)\cos(\beta)$) and arc orientations $\delta$, and therefore their changes $\varepsilon$ ($\varepsilon_i = \delta_i - \delta'_{i-1}$) are independent random variables. The distribution of the former is therefore characterised by a mean cosine $c_\omega = c_\alpha c_\beta$, and the distributions of the later are centrally symmetrical, i.e. characterised by null mean cosines and sines ($c_\delta = c_\varepsilon = s_\delta = s_\varepsilon = 0$).

By applying the spherical law of cosines to the three orthodromic arcs linking any three successive 3D headings ($\theta_{i-1}$, $\phi_{i-1}$), ($\theta_i$, $\phi_i$) and ($\theta_{i+1}$, $\phi_{i+1}$), and whose sizes are $\Omega_{i-1,i} = \omega_{i-1}$, $\Omega_{i,i+1} = \omega_i$ and $\Omega_{i-1,i+1}$, respectively, one gets:

$$\cos(\Omega_{i-1,i+1}) = \cos(\omega_{i-1})\cos(\omega_i) - \sin(\omega_{i-1})\sin(\omega_i)\cos(\varepsilon_i).$$

Call $\Omega_k$ the size of the orthodromic arc corresponding to any $k$-order reorientation, i.e. the angular deviation between any couple of steps that are separated by $k-1$ steps ($\Omega_1 = \omega$). For any three successive steps, as $\omega$ and $\varepsilon$ are independent random variables, and the latter is centrally symmetrical ($c_\varepsilon = s_\varepsilon = 0$), one gets $E[\cos(\Omega_2)] = c_\omega^2$, and by extension $E[\cos(\Omega_k)] = c_\omega^k$ for any $k$. Given that $\Omega_k$ appears $n-k$ times in an $n$-step path, the MSD of a 3D CRW can therefore be expressed as:

$$E(R_{n,3D}^2) = nE(l^2) + 2E(l)^2 \sum_{k=1}^{n-1}(n-k)c_\omega^k \tag{3}$$

and therefore (based on the geometric series property $\sum_{i=p}^{q} x^i = (x^p - x^{q+1})/(1-x)$ for $p \geq q$):

$$E(R_{n,3D}^2) = nE(l^2) + E(l)^2 \frac{2c_\omega}{1-c_\omega}\left(n - \frac{1-c_\omega^n}{1-c_\omega}\right) \tag{4}$$

An equivalent formulation in terms of yaw and pitch is obtained by replacing $c_\omega$ by the product $c_\alpha c_\beta$. Eq. (4) generalises in 3D the formulation for 2D balanced CRW, initially derived by Tchen (1952) and Skellam (1973) for a constant step length ($E(l^2) = E(l)^2$) and expanded by Hall (1977) when the step length is an independent random variable ($E(l^2) > E(l)^2$), which is obtained simply by replacing $c_\omega$ by the mean cosine of 2D rotations. This means that the MSD does not depend on whether turning angles are all performed in the same plane (2D movements) or performed in a new plane (rotated by $\xi$ with respect to the previous one) at each step (3D movements).

When the step number $n$ is large, Eq. (4) can be approximated as:

$$E(R_{n,3D}^2)_a = nE(l)^2\left(\frac{1+c_\omega}{1-c_\omega} + b^2\right) = nE(l)^2\left(\frac{1+c_\alpha c_\beta}{1-c_\alpha c_\beta} + b^2\right) \tag{5}$$

where $b$ is the coefficient of variation of step length ($E(l^2) = E(l)^2(1+b^2)$). Eq. (5) shows that a CRW, in 3D as in 2D, tends to be diffusive in the long term: the MSD is almost proportional to the step number $n$, i.e. the path length, when $n$ is large.

The MSD of a diffusive walk after a time $T$ is equal to $2(D_x+D_y+D_z)T$, where $D_x$, $D_y$ and $D_z$ are the diffusion coefficients for $X$, $Y$ and $Z$ axes (Codling et al., 2008), and therefore $4DT$ for a 2D isotropic diffusive walk ($D_x = D_y = D$; $D_z = 0$) and $6DT$ for a 3D isotropic diffusive walk ($D_x = D_y = D_z = D$). The sinuosity of 2D balanced CRW was defined by Benhamou (2004) as $S_{2D} = [4L/E(R_{n,2D}^2)_a]^{0.5}$, where $L = nE(l)$ is the mean path length after $n$ steps. The coefficient of diffusion can therefore be expressed as $D = V/S_{2D}^2$ (Benhamou, 2006). In other terms, the sinuosity corresponds to the purely spatial component of an isotropic diffusion process. To keep the same kind of relationship in 3D, the sinuosity can be defined as $S_{3D} = [6L/E(R^2_{n,3D})_a]^{0.5}$. This leads to the following expression when reorientations are expressed in terms of orthodromic arc sizes ω, or in terms of yaw α and pitch β:

$$S_{3D} = \left[\frac{E(l)}{6}\left(\frac{1+c_\omega}{1-c_\omega}+b^2\right)\right]^{-0.5} = \left[\frac{E(l)}{6}\left(\frac{1+c_\alpha c_\beta}{1-c_\alpha c_\beta}+b^2\right)\right]^{-0.5} \quad (6)$$

When analysing actual 3D movements, it is therefore important to check that the distribution of changes in arc orientations ε is centrally symmetrical ($c_\varepsilon = s_\varepsilon = 0$), and also that step lengths *l*, orthodromic arc sizes ω and initial orientations δ are independent random variables (neither autocorrelated nor cross-correlated). Otherwise the actual 3D movement cannot be assimilated to a balanced CRW and Eq. (6) does not apply. Although they are useful to model random search paths, 2D or 3D CRWs cannot be expected to realistically represent most actual movements. CRW nevertheless act as useful null models to highlight more sophisticated movement behaviours. This has been the case for a long time in 2D. With the current increasing number of studies on 3D movements, the MSD and sinuosity expressions developed here should help analysing them.

arXiv:1801.02435

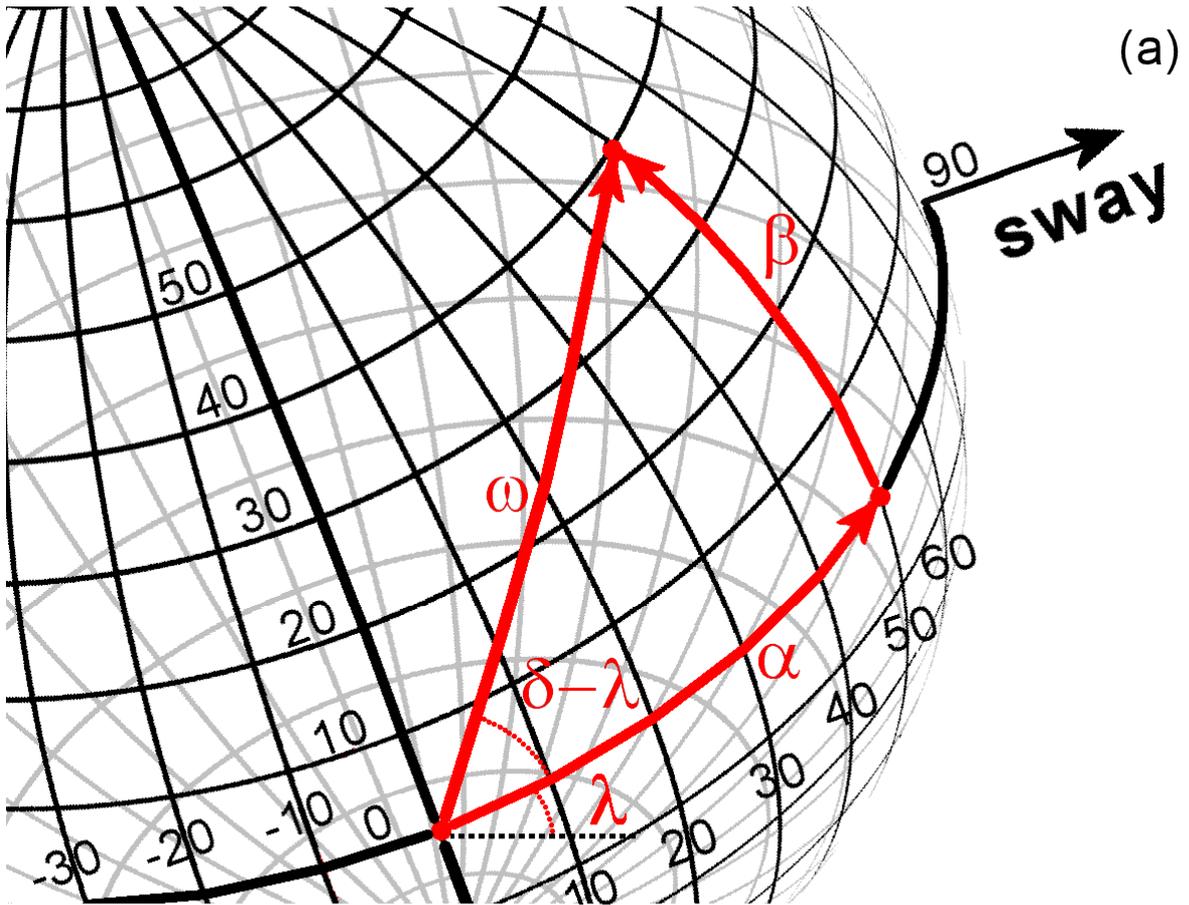

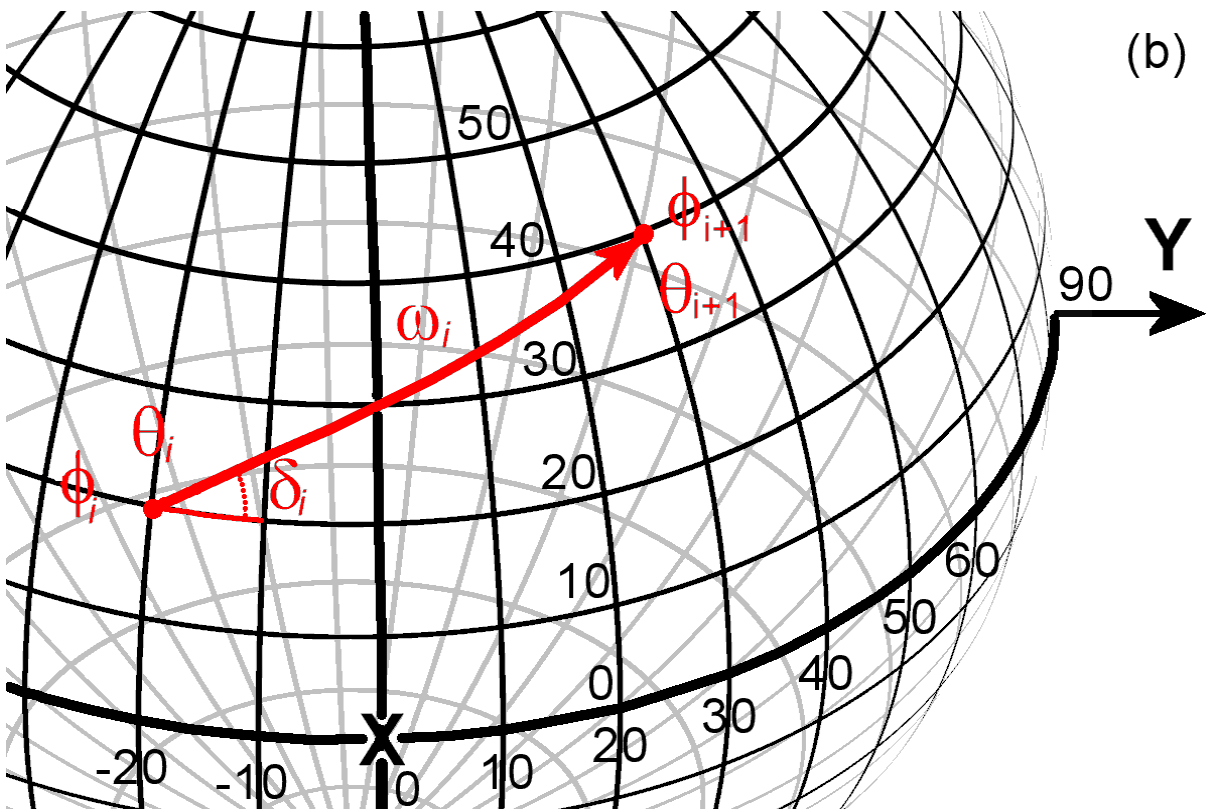

**Fig. 1. Relationships between the various angles on the unit sphere**.

(a) Reorientation in the egocentric frame of reference (surge, sway and heave axes) in terms of yaw, $\alpha$, and pitch, $\beta$, performed respectively in the coronal (or "equatorial"; surge axis x sway axis) plane, and orthogonally to it in the sagittal (surge axis x heave axis) plane, after this plane has been rotated by $\alpha$. The origin at the sphere surface (0, 0) corresponds to the initial heading. Given initial lateral inclination $\lambda$, this reorientation can be expressed in terms of angular length, $\omega$, and initial orientation in the frontal plane, $\delta$, of the orthodromic arc. In this example angular values are: $\alpha = 50°$, $\beta = 40°$, $\lambda = 20°$, $\omega = 60.5°$, $\delta = 67.6°$.

(b) Reorientation in an exocentric frame of reference (*X*, *Y*, *Z*) from heading ($\theta_i$, $\phi_i$) to heading ($\theta_{i+1}$, $\phi_{i+1}$), expressed in terms of angular length, $\omega_i$, and initial orientation, $\delta_i$, of the orthodromic arc. In the present context, $\theta$ and $\phi$ correspond to the azimuth and the longitudinal inclination of the animal's body, but may be interpreted as longitude and latitude, respectively, in the context of 2D navigation on the Earth's surface. In this example, angular values are: $\theta_i = -20°$, $\phi_i = 20°$, $\theta_{i+1} = 30°$, $\phi_{i+1} = 40°$, $\omega_i = 47.0°$, $\delta_i = 36.6°$.

The spherical wireframe used as background come from

https://en.wikipedia.org/wiki/Sphere#/media/File:Sphere_wireframe_10deg_6r.svg.

**APPENDIX: SPHERICAL TRIGONOMETRIC RELATIONSHIPS**

The following expressions were derived using the so-called East-North-Up frame of reference in which angles are measured in agreement with trigonometric conventions. The expressions linking ($\omega$, $\delta$) and ($\theta$, $\phi$) provided here may differ from those classically used in great-circle navigation (where $\theta$ and $\phi$ correspond to longitude and latitude, respectively) because of different conventions (in navigation studies, directions are often expressed in the so-called North-East-Down frame of reference, involving that angles in the horizontal plane are measured clockwise from North rather than anticlockwise from an *X* axis pointing to East). Note: The atan$_2$ function, which appears in several expressions below, corresponds to four-quadrant arctangent: atan$_2$($y$, $x$) = arctan($y/x$) for $x>0$ and atan$_2$($y$, $x$) = arctan($y/x$) $\pm$ $\pi$ for $x<0$.

Given a couple of successive locations $\mathbf{X}_{i-1} = (X_{i-1}, Y_{i-1}, Z_{i-1})$ and $\mathbf{X}_i = (X_i, Y_i, Z_i)$, with $X_i = X_{i-1} + l_i \cos(\phi_i) \cos(\theta_i)$, $Y_i = Y_{i-1} + l_i \cos(\phi_i) \sin(\theta_i)$, and $Z_i = Z_{i-1} + l_i \sin(\phi_i)$, azimuth $\theta_i$ and longitudinal inclination $\phi_i$, are given by:

$$\theta_i = \text{atan}_2(Y_i - Y_{i-1}, X_i - X_{i-1})$$

$$\phi_i = \sin^{-1}[(Z_i - Z_{i-1})/l_i]$$

Based on successive rotations of *X*, *Y* and *Z* axes, it can be shown that the cosines and sines of the angular size $\Omega_{i,j}$ and initial orientation $\Delta_{i,j}$ of any orthodromic arc linking any two 3D headings ($\theta_i$, $\phi_i$) and ($\theta_j$, $\phi_j$) are related to the cosines and sines of these headings as follows:

$$\cos(\phi_j) \cos(\theta_j - \theta_i) = \cos(\Omega_{i,j}) \cos(\phi_i) - \sin(\Delta_{i,j}) \sin(\Omega_{i,j}) \sin(\phi_i)$$

$$\cos(\phi_j) \sin(\theta_j - \theta_i) = \cos(\Delta_{i,j}) \sin(\Omega_{i,j})$$

$$\sin(\phi_j) = \sin(\Delta_{i,j}) \sin(\Omega_{i,j}) \cos(\phi_i) + \cos(\Omega_{i,j}) \sin(\phi_i)$$

By recombining these expressions, it can be shown that the reorientation ($\omega_i = \Omega_{i,i+1}$, $\delta_i = \Delta_{i,i+1}$) between two successive headings ($\theta_i$, $\phi_i$) and ($\theta_{i+1}$, $\phi_{i+1}$) is given by:

$$\omega_i = \cos^{-1}[\sin(\phi_i) \sin(\phi_{i+1}) + \cos(\phi_i) \cos(\phi_{i+1}) \cos(\theta_{i+1} - \theta_i)],$$

$\delta_i = \text{atan}_2[\sin(\phi_{i+1}) - \cos(\omega_i) \sin(\phi_i), \cos(\phi_i) \cos(\phi_{i+1}) \sin(\theta_{i+1}-\theta_i)]$

and symmetrically: $\delta'_i = \text{atan}_2[\cos(\omega_i) \sin(\phi_{i+1}) - \sin(\phi_i), \cos(\phi_i) \cos(\phi_{i+1}) \sin(\theta_{i+1}-\theta_i)]$.

For a given initial lateral inclination $\lambda_i$, any reorientation expressed in terms of $(\omega_i, \delta_i)$ can be translated in terms of $(\alpha_i, \beta_i)$ and vice-versa, using the following basic relationships, derived for simplicity by considering that yaw $\alpha_i$, pitch $\beta_i$ and roll $\rho_i$ occur successively in this order rather than simultaneously:

$\cos(\omega_i) = \cos(\alpha_i) \cos(\beta_i)$

$\sin(\beta_i) = \sin(\omega_i) \sin(\eta_i)$

$\sin(\alpha_i) \cos(\beta_i) = \sin(\omega_i) \cos(\eta_i)$

where $\eta_i = \delta_i - \lambda_i = \delta_i' - \lambda_i'$ corresponds to the inclination of the rotation plane (as defined by the orthodromic arc) with respect to the coronal plane, and controls how the reorientation is shared between yaw and pitch components, from pure yaw ($\eta_i = 0$ or $\eta_i = \pm\pi$) to pure pitch ($\eta_i = \pm\pi/2$).

Reciprocally, knowing reorientation expressed in terms of orthodromic arc $(\omega_i, \delta_i)$ or in terms of yaw, pitch and roll $(\alpha_i, \beta_i, \rho_i)$, the new 3D orientation $(\theta_{i+1}, \phi_{i+1}, \lambda_{i+1})$ can be determined from the previous one $(\theta_i, \phi_i, \lambda_i)$ as follows:

$\theta_{i+1} = \theta_i + \text{atan}_2[\sin(\omega_i) \cos(\delta_i), \cos(\omega_i) \cos(\phi_i) - \sin(\omega_i) \sin(\delta_i) \sin(\phi_i)]$

$\qquad = \theta_i + \text{atan}_2[\sin(\alpha_i) \cos(\beta_i) \cos(\lambda_i) - \sin(\beta_i) \sin(\lambda_i),$

$\qquad\qquad \cos(\alpha_i) \cos(\beta_i) \cos(\phi_i) - (\sin(\beta_i) \cos(\lambda_i) + \sin(\alpha_i) \cos(\beta_i) \sin(\lambda_i)) \sin(\phi_i)]$

$\phi_{i+1} = \sin^{-1}[\sin(\omega_i) \sin(\delta_i) \cos(\phi_i) + \cos(\omega_i) \sin(\phi_i)]$

$\qquad = \sin^{-1}[(\sin(\beta_i) \cos(\lambda_i) + \sin(\alpha_i) \cos(\beta_i) \sin(\lambda_i)) \cos(\phi_i) + \cos(\alpha_i) \cos(\beta_i) \sin(\phi_i)]$

$\lambda_{i+1} = \lambda_i + \rho_i + \delta'_i - \delta_i$